# Tunable multi-color coherent light generation in single MgO: PPLN bulk crystal.

CHOGE DISMAS K.[1,2], CHEN HUAI-XI[1], LEI GUO[1], XU YI-BIN[1], LI GUANG-WEI[1], AND LIANG WAN-GUO[1,*]

[1]*Fujian Institute of Research on the Structure of Matter, Chinese Academy of Sciences, Fuzhou, 350002, PR China*

[2]*University of Eldoret, Eldoret, 30100, Kenya*

*\* wgl@fjirsm.ac.cn*

**Abstract:** We report a theoretical design analysis of domain-engineered periodically poled lithium niobate (PPLN) for wavelength conversion of near-infrared sources to generate coherent light in the visible spectral range. Our analysis on the spectral outputs show that with a proper design of the quasi phase matching (QPM) periods, tunable, multiple nonlinear optical processes can be simultaneously phase matched in a single segmented crystal. We show that a three-segment single PPLN crystal has potential to generate violet (432 nm), blue (490 nm) and orange (600 nm) wavelengths by simultaneous sum frequency and second harmonic generation processes. Such a design scheme has promising potential for a compact, robust and tunable multi-colored visible light source which can find various applications such as in biomedicine, high-density optical data storage and laser based color displays.

1. Introduction

Quasi phase-matching (QPM) in nonlinear optical crystals such as lithium niobate (LN) has attracted significant attention over the last 3 decades primarily due to the flexibility it offers in engineering any desired phase matching condition by spatially modulating the quadratic nonlinearity[1]. QPM second harmonic generation (SHG) and sum frequency generation (SFG) are particularly attractive techniques for generating visible light. On the other hand, multi-periodic poling of LN offers an advantage of achieving simultaneous processes within the same crystal. By using this technique, simultaneous generation of multiple wavelengths in the visible range within the same crystal have been realized which has found use in many applications such biomedicine, high-density optical data storage and laser based color displays. Previously, simultaneous generation of the three primary colors: red (R), green (G) and blue (B) has been demonstrated. For instance, Brunner *et al.* reported RGB source by using several crystals each for different nonlinear process[2]. Capmany and co-workers demonstrated RGB source and continuous wave (CW) green and blue lasers using self-frequency doubling and self-frequency mixing in $Nd^{3+}$ -doped bulk aperiodically poled lithium niobate in[3, 4]. Liao *et al.* reported red, yellow, green and blue light in a single aperiodically poled lithium tantalate (LT) [5, 6]. Lim *et al.* demonstrated simultaneous RGB generation in PPLN with ultra-broadband optical parametric generation (OPG)[7] while Gao and Xu *et al.* demonstrated RGB laser generation by cascaded nonlinear interactions in a single stoichiometric LT with two periodicities[8, 9].

Our aim here is to investigate single-pass SFG and SHG phase matching processes in a three-segment QPM grating structure to simultaneously generate CW visible light at multiple wavelengths. This scheme has potential for a compact, robust and tunable multi-colored source of visible light within the same crystal.

### 2. PPLN design

The QPM grating structure shown in Figure 1 is assumed to be implemented in a z-cut x-propagating magnesium oxide doped periodically poled lithium niobate (MgO: PPLN) bulk crystal and consists of three segments with total device length L. The periods are denoted as $\Lambda_1$, $\Lambda_2$ and $\Lambda_3$ for segment 1, 2 and 3 respectively. For the first two segments intended for SFG, the coupled mode equations are expressed as[10]

$$\frac{dA_{1,m}}{dz} = -\frac{2i\omega_{1,m}}{n_{1,m}c} d_{eff} A_{3,m} A^*_{2,m} \exp(-i\Delta k_m z) \tag{1}$$

$$\frac{dA_{2,m}}{dz} = -\frac{2i\omega_{2,m}}{n_{2,m}c} d_{eff} A_{3,m} A^*_{1,m} \exp(-i\Delta k_m z) \tag{2}$$

$$\frac{dA_{3,m}}{dz} = -\frac{2i\omega_{3,m}}{n_{3,m}c} d_{eff} A_{1,m} A_{2,m} \exp(i\Delta k_m z) \tag{3}$$

Where $A_{1,m}$, $A_{2,m}$, and $A_{3,m}$ denote the amplitudes of the pump ($\omega_{1,m}$), signal ($\omega_{2,m}$), and idler ($\omega_{3,m}$), respectively, c is the speed of light in vacuum and *z* is the distance along the direction of propagation. The subscript m denotes either the first or the second segment. Here, the phase mismatch due to material dispersion is expressed as [11]

$$\Delta k_m = k_{1,m} + k_{2,m} - k_{3,m} - \frac{2\pi}{\Lambda_m} \tag{4}$$

The temperature dependent refractive indices $n_{1,m}$, $n_{2,m}$ and $n_{3,m}$ are calculated using the Sellmeier equations, and $d_{eff} = 2d_{33}/\pi$ ($d_{33}$= 27 pm/V) is the effective nonlinear coefficient of MgO: PPLN [12]. This highest nonlinearity can be accessed when all the three interacting waves are polarized along the optical axis of the MgO: PPLN crystal. For SHG in the third segment, the coupled mode equations take the form

$$\frac{dA_2}{dz} = -\frac{2i\omega_2}{n_2 c} d_{eff} A_5 A^*_2 \exp(-i\Delta kz) \tag{5}$$

$$\frac{dA_5}{dz} = -\frac{2i\omega_5}{n_5 c} d_{eff} A_2^2 \exp(i\Delta kz) \tag{6}$$

and the phase mismatch is

$$\Delta k = k_5 + 2k_2 - \frac{2\pi}{\Lambda_3}. \quad (7)$$

Where $A_5$ is the amplitude of the second harmonic of $\omega_2$ at $\omega_5$. The normalized conversion efficiencies for SFG and SHG processes are estimated as the power ratio between the output SFG/SHG power and the product of the input power, which are expressed respectively as

$$\eta_{SFG} = P_{SFG} / (P_1 P_2) W^{-1} \quad (8)$$

$$\eta_{SHG} = P_{SHG} / (P_2)^2 W^{-1} \quad (9)$$

For a single-pass SFG and SHG configurations, Eqs. (1)– (3) and Eqs. (5)- (6) are simultaneously solved assuming undepleted pump regime and negligible propagation losses.

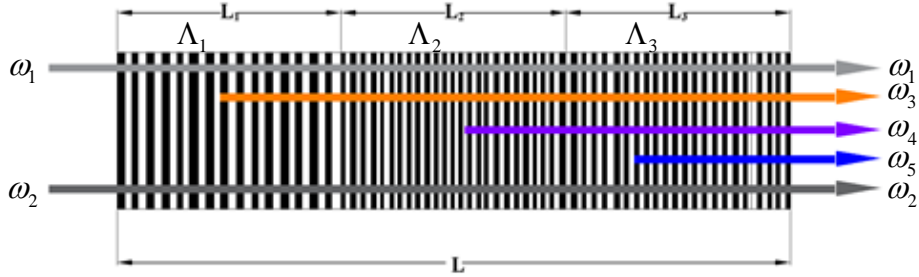

**Figure 1. Three segment grating structure**

Table 1. Summary of the device design parameters.

| Process | $\Lambda_{QPM}$ ($\mu m$) at 25 °C [a] | Segment Length (mm) | Expected output wavelength (nm) |
|---|---|---|---|
| SFG($\omega_1 + \omega_2 \rightarrow \omega_3$) | 10.4 | 15 | 600 |
| SFG($\omega_1 + \omega_3 \rightarrow \omega_4$) | 4.3 | 15 | 432 |
| SHG ($\omega_2 \rightarrow \omega_5$) | 5.4 | 15 | 490 |

[a] calculated with SNLO (AS photonics)

A tunable pump ($P_1$=2 W) in the wavelength range from 1500 nm to 1600 nm is supposedly injected in to the PPLN along with a fixed signal ($P_2$=1 W) at central wavelength of 980 nm. As the pump and the signal propagate through the first segment of PPLN crystal, it generates an idler light in the 600 nm wavelength range by sum frequency interaction. The sum frequency idler then interacts (sum frequency interaction) with the pump in the second segment of the crystal leading to generation of a second idler in the 432 nm wavelength range. The 490 nm wavelength output is generated by frequency doubling of the residual signal in the last segment of the crystal. The simultaneous processes are dependent on the temperature of the crystal, phase mismatch between the pump, signal and the two idlers while the conversion efficiency depends on the proper phase matching and the fundamental powers.

### 3. Results and discussions

We simulated the SFG and SHG processes by solving the coupled mode equations using the QPM periods predicted with SNLO (AS photonics) for MgO: PPLN (see Table 1). We assumed that both the pump and the signal are highly monochromatic and undepleted. The powers of the input light sources are assumed to be 2 W and 1 W for the pump and signal respectively and output power of idler 1 acts as the signal for the second segment in the calculation. The wavelength of the tunable pump laser was increased in steps of 0.001 nm. For comparison, we assume that the three segments are of equal length. Figure 2 shows the simulated conversion efficiencies of both idler 1 and 2 generated by sum frequency interaction in segment 1 and 2 and second harmonic along with pump and signal wavelengths. The corresponding idler and second harmonic wavelengths are shown in the upper horizontal axes. Here, the PPLN temperature is increased in steps of 1 $^oC$ from 25 $^oC$ to 31 $^oC$. We notice that the SFG process in the second segment of the device is very sensitive to slight temperature change with its conversion efficiency decreasing rapidly with increase in temperature (Figure 2 b). However, the SFG process in the first segment has a broad temperature tolerance (Figure 2 a) while the SHG process is slightly sensitive to temperature. Therefore, it is possible to access up to 6.0 nm in the 600 spectral region by simply changing the temperature as shown in figure 2(a). Similarly, up to 7.1 nm in the 432 nm region and up to 3.9 nm in the 490 nm region as shown in figures 2 (b-c). For the results shown in figure 2 (c), the signal is tuned from 965 nm to 1000 nm in steps of 0.001 nm. According to figure 2 (b), we can also notice the dual peaks at 434.9 nm and 435.4 nm for 27 $^oC$ and, 435.9 nm and 439.9 nm for 28 $^oC$.

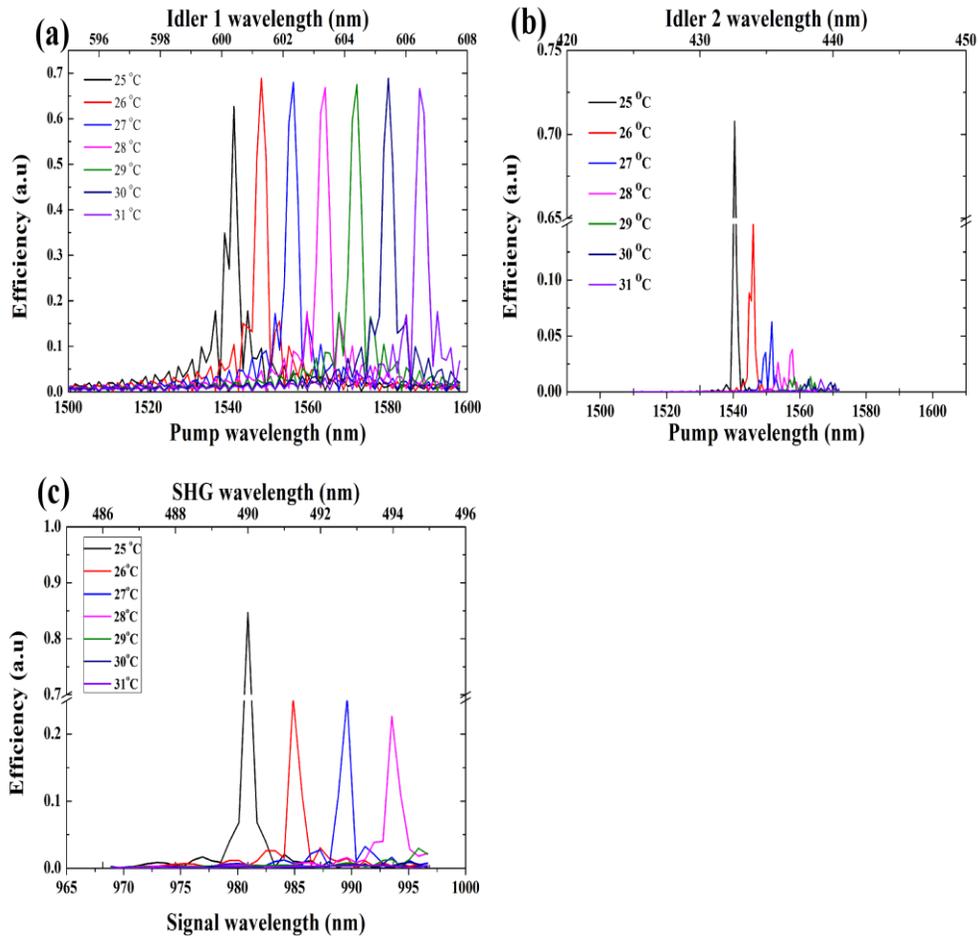

Figure 2. Efficiency as a function of pump and signal wavelengths at various temperatures for (a) 600 nm, (b) 432 nm and (c) 490 nm spectral regions

Figure 3 shows the overlaying of simulated spectra for SFG 1, SFG 2 and SHG for two different pumps centered at 1553 nm and 1554 nm and the signal at 980 nm when the temperature is fixed at 25 °C. It can be seen that the conversion efficiencies for SFG 1, SFG 2 and SHG are approximately 62%, 65% and 85% respectively.

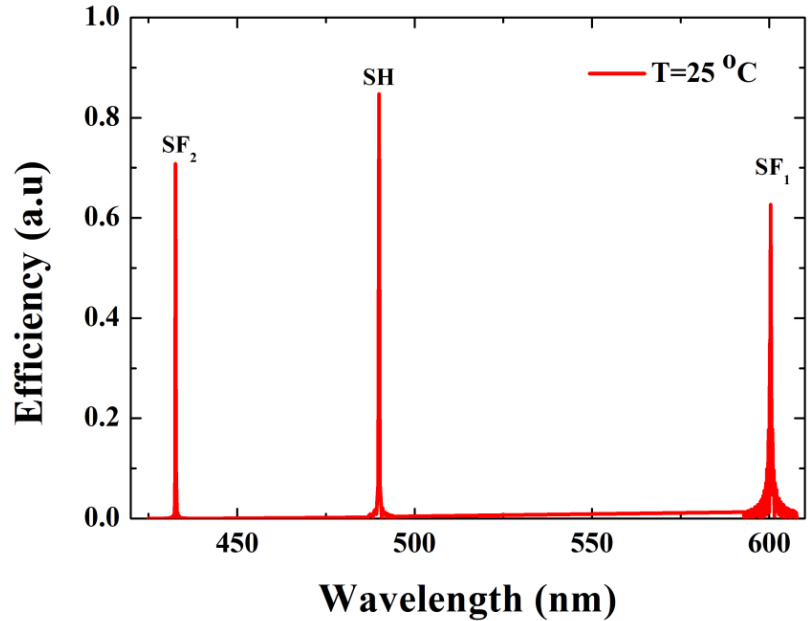

**Figure 3.** Overlaying of three simulated spectra including SFG of two pumps at 1553 nm and 1554 nm and SHG of signal at 980nm named as SF for the pumps and the signal presented as SH.

Therefore by tuning the pump to 1553 nm and 1554 nm, both SFG processes in the first two segments are satisfied respectively for a fixed signal with comparative conversion efficiencies and at the same time, an efficient SHG at 490 nm is obtained in the third segment.

Figure 4 shows the calculated wave vectors for SFG 1, SFG 2 and SHG in segments 1, 2 and 3 respectively. The results presented here are for central wavelengths of 1550 nm, 980 nm and 600 nm for pump, signal and idler 1 respectively. Here, we used the numerical method proposed in Ref. [13] where the beam waist of 100 µm was assumed for input lasers although this is not the optimized value. At the end of each section, $K_x$ and $K_y$ wave vector components are seen to be wider compared to the fundamental beams in the spatial spectra which is expected from wavelength differences.

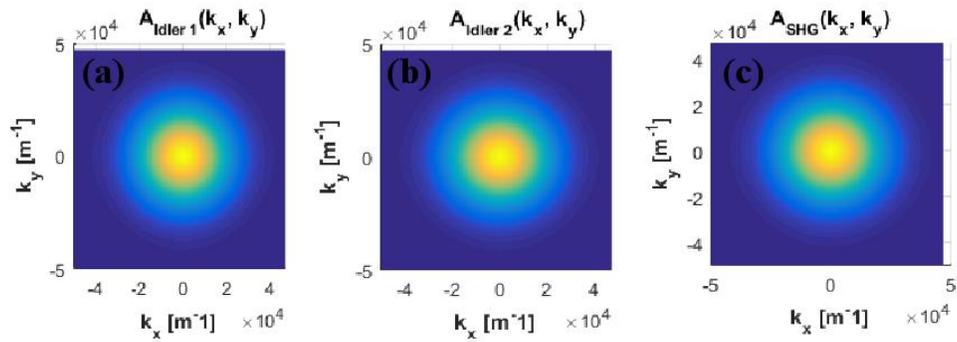

**Figure 4.** Wave vector profiles for the simulated outputs at (a) 600 nm, (b) 432 nm and (c) 490 nm spectral regions.

4. **Conclusion**

We have reported a theoretical design analysis of QPM segmented gratings in PPLN for wavelength conversion of near-infrared laser sources to generate coherent light in the visible spectral range by simultaneous SFG and SHG processes within a single crystal. The analysis on the spectral outputs show possible temperature tuning to access up to 6.0 nm, 7.1 nm and 3.9 nm in the 600 nm, 432 nm and 490 nm spectral regions respectively. Such design scheme offers a promising potential for a compact, robust and tunable multi-colored source of visible light within the same crystal which can find applications in biomedicine, high-density optical data storage and laser-based color displays.

**Funding**

Fujian Science and Technology Service Network Initiative (STSI) Project (2016T3010).